\documentstyle[aps,prl,multicol,epsf]{revtex}
\epsfclipon

\draft

\begin{document}
\author{V. N. Bogomolov}
\address{A. F. Ioffe Physicotechnical Institute, Russian Academy of Sciences, 194021\\
St. Petersburg, Russia}
\title{Metallic Xenon, Molecular Condensates, and Superconductivity}
\date{\today}
\maketitle

\begin{abstract}
A possibility of explaining the light absorption observed to occur under
pressure-induced xenon metallization as due to the transition to the
superconducting state is analyzed. The mechanism of the van der Waals
bonding is discussed.
\end{abstract}


\begin{multicols}{2}

\narrowtext

---{\em Introduction. }The evolution of the physical model underlying the
theory of superconductivity of metals (the BCS theory) is well known. The
history of investigating the possibility of metallization of molecular
condensates (MC) and the onset of superconductivity in these compounds dates
back much farther. In the XVIII century, it was assumed that hydrogen in the
condensed state may be a metal \cite{1}. This problem remains unsolved up to
now \cite{2}. A simple criterion of MC metallization under pressure was
found in 1927 \cite{3}. The conditions favorable for the metallization of
atomic hydrogen were theoretically estimated in 1935 \cite{4}. In 1938, the
superconductivity was identified with boson condensation \cite{5}. In 1946,
the first attempts at condensing bosons as electron pairs in bubbles,
supposed to form when sodium dissolves in ammonia, were made \cite{6}. One
observed induced currents, which decayed extremely weakly at $80K$. These
results, however, were subsequently questioned \cite{7}. In 1949,
theoretical and experimental investigation of the processes initiated by
bringing atoms closer together were started \cite{8,9}. The interest in the
problem of metallic hydrogen and progress in the high static-pressure
technologies have culminated in 1989 in a study by optical means of the
metallization of xenon at pressures from $130$ to $200GPa$ \cite{10}. The
data obtained were interpreted as indicating band-gap closure induced by
lattice compression. In 2000, the dc conductivity of $Xe$ was studied up to $%
155GPa$ \cite{11}.

---{\em Analysis of the experimental data of Goettel et al. }\cite{10}. A
reliable proof of the onset of superconductivity can be obtained only by
magnetic measurements. Optical data are the least appropriate for this
purpose. Nevertheless, the extremely high accuracy of the data of \cite{10}
and, at the same time, their significant deviation from band theory
predictions justify an attempt at their interpretation within a different
framework \cite{12}.

\begin{figure}
\epsfxsize=80mm
\epsffile{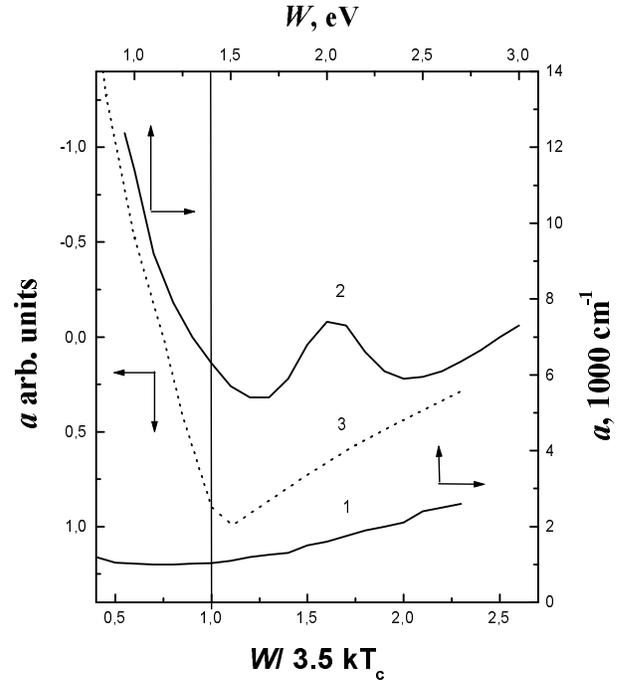}
\caption{
Light absorption coefficient $a$ in xenon plotted against photon
energy \protect\cite{10}. Pressure ($GPa$): (1) - $130$, (2) - $200$.
Dashed curve: absorption coefficient of superconducting films calculated
vs. $W/kT_c$, the ratio of photon to gap energy \protect\cite{13}.
}
\label{fig1}
\end{figure}

Figure \ref{fig1} presents the dependences of $a$, the light absorption
coefficient, on photon energy $W$ in a xenon film calculated from the data
of \cite{10} for the pressures of 130 GPa (curve 1) and 200 GPa (curve 2).
Curve 3 plots the coefficient $a$ calculated for superconductor films as a
function of $W/(3.5kT_c)$, i.e., the ratio of photon energy to the
superconducting transition temperature \cite{13}. The right-hand part of the
curves can be explained as due to absorption by normal electrons. Curves of
type 3 exhibit sometimes a precursor hump, similar to curve 2 for xenon. A
comparison of curves 2 and 3 permits one to estimate formally the $T_c$
point of xenon at $200GPa$ as about $4000K$.

By the band model of metallization, the plasma frequency $W_p^{4/3}$ and the
total absorption $b^{2/3}$ in the region of the peak should be proportional
to the molar volume difference from the volume at metallization,
i.e., $\sim
(V_m-V)$ \cite{10}. These dependences were used to determine \cite{10} $%
V_m=10.7-10.5cm^3/mol$ and the pressure $P_m=130-140GPa$. Figure \ref{fig2}
plots $W_p^2$ and $b^1$ vs. $(V_m-V)$, i.e., they are the same dependences
but with other exponents. These functions approximate better the
experimental points of \cite{10}, which permits the following two
conclusions:

\begin{figure}
\epsfxsize=85mm
\epsffile{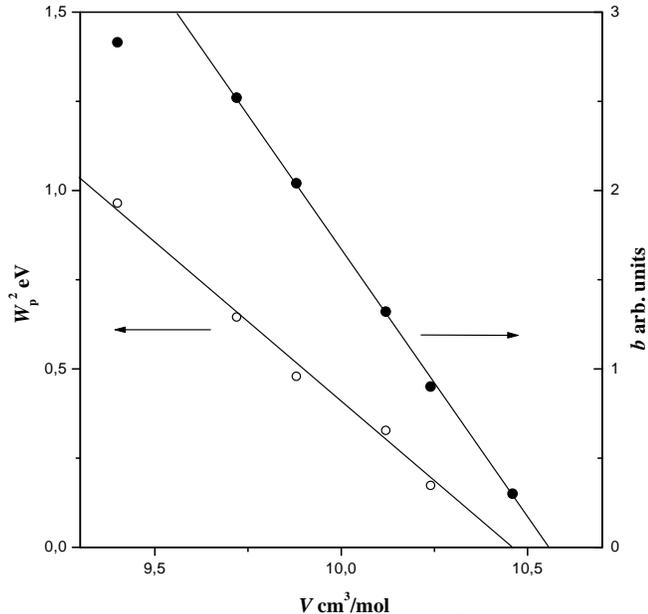}
\caption{
Squared plasma frequency $W_p^2$ and total absorption coefficient $b$
plotted vs. xenon volume $V$ (data of \protect\cite{10}).}
\label{fig2}
\end{figure}

(i) The first function can be approximated by:

\[
0.5W_p=3.2B(1-V/V_m)^{1/2}\;\text{for }B=0.48eV\,.
\]
This relation can describe the gap energy $W_p$ in the region of a
phase transition, for instance, to the superconducting state. The Gibbs
function depends on the parameters $T$ and $V$ in the same way. In this
case, $B=T_m=T_c=5000K$, which is close to the above estimate, and the
gap energy is of the order of $1\div 1.5eV$.

\begin{figure}
\epsfxsize=85mm
\epsffile{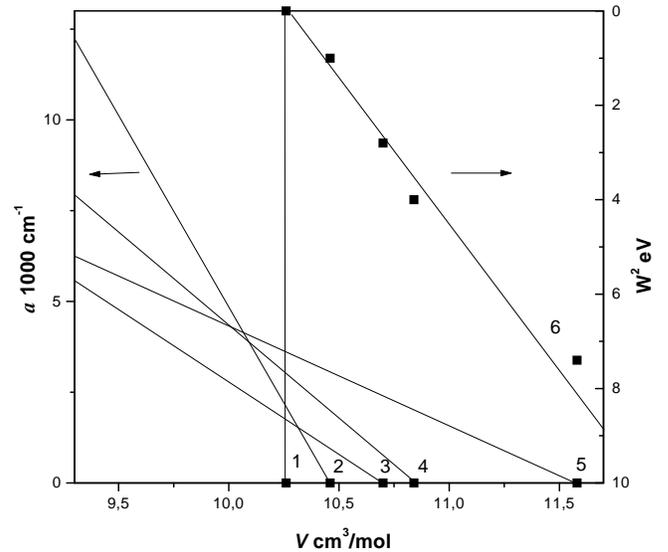}
\caption{
Absorption coefficient $a$ plotted vs. xenon volume $V$ for energies
($eV$): (1) - 0, (2) - 1.0, (3) - 1.7, (4) - 2.0, (5) - 2.7
(data of \protect\cite{10}). The onset
of metallization depends on the light frequency. Curve (6):
dependence of the
volume at metallization on squared photon energy $W^2$.
}
\label{fig3}
\end{figure}

(ii) The transition of xenon to the superconducting state can be described
also in terms of the percolation process \cite{14}. In this case, $V_m$ and $%
P_m$ should depend on photon energy $W$. In Fig. \ref{fig3}, the data of
\cite{10} were used to plot the dependence of the absorption coefficient on
xenon volume for different photon energies. The volume at which absorption
appears ($a\sim 0$) depends on frequency. The dependence of these volumes on
$W^2$ makes it possible to determine the onset of metallization in the case
of dc measurements: $V_{m0}=10.26cm^3/mol$ and $P_{m0}=152GPa$.
By the Herzfeld
criterion, $V_{m0}=10.20cm^3/mol$ and $P_{m0}=154GPa$ \cite{10}. Therefore
in the dc conductivity measurements carried out in \cite{11} up to $155GPa$,
the xenon metallization conditions were practically not reached, and the
weak growth of the resistance with temperature is possibly due to the effect
of contacts and lattice vibration anharmonicity on the total sample
resistance.

---{\em Van der Waals bonding.} In describing MCs one conventionally makes
use of various kinds of averaging (for instance, by introducing the
Lennard--Jones potential). This approach may result in either loosing or
wrongly interpreting a number of their properties. Consider a condensate of $%
Xe$ atoms ($5s^25p^6$). We assume the xenon atoms in a virtually excited
state to have the ($5s^25p^56s$) configuration. In this case, $(Xe)$ is a
chemical analog of $Cs$ ($5s^25p^66s$). The interatomic distances in solid $%
Xe$ and in $Cs$ are similar ($4.4$ and $4.6\AA $). The hydrogenlike radius
of ground-state $Xe$, $r_1=e^2/2E_1=0.59\AA $ ($E_1=12.13eV$ is the
ionization potential), and the excited-state radius $r_2=e^2/2(E_1-E_2)=1.88%
\AA $ ($E_2=8.3eV$ is the transition energy to the excited state). In
condensates, the atomic radii increase by $15-20\%$, which yields $r_2=2.2%
\AA $. (The effect reverse to the Goldschmidt compression accompanying the
removal of an atom from the condensate.) That the interatomic radii in
molecular condensates are determined by the excited-state radii, i.e., by
the $(E_1-E_2)$ energies, is well known. However, the physical
interpretation of this situation may be ambiguous. The distances between
ground-state atoms are large ($r_2/r_1=3.7>1$), and the time-averaged
population of the excited-state orbitals $X\ll 1$. The transition
probability is low, as is the exciton concentration. Therefore, as in the
case of small-radius polarons, the electronic states may be considered well
localized, thus permitting one to abandon the band models.

The ground-state wave functions of neighboring atoms in the xenon
condensate, for instance, $F_{5s,5p}\symbol{126}\exp \left( -r/5r_1\right) $%
, overlap only weakly. The prefactor is dropped, and the Bohr radius of $0.53%
\AA $ is replaced by $r_1=0.59\AA $. The wave functions of neighboring atoms
coupled by van der Waals bonding are not correlated (unlike the covalent
bonding). Therefore the probability for two electrons of neighboring atoms
to occupy simultaneously the same point $r$, to become excited, and to
transfer to the excited-state orbitals can be written as:
\begin{eqnarray*}
X &\sim &\exp \left( -2\frac r{5r_1}\right) \exp \left( -2\frac{2r_2-r}{5r_1}%
\right) =\exp \left( -\frac{4r_2}{5r_1}\right) \,, \\
X &\sim &\exp \left( -0.8\frac{E_1}w\right) \;\text{for \ }w=\frac{e^2}{2r_2}%
\,.
\end{eqnarray*}
This relation may be regarded as the probability to overcome a barrier $%
0.8E_1$ when acted upon by a random perturbation of average energy $w$. The
energy of an atomic transition to the excited state $E_2$ ($\sim 0.7E_1$ for
Xe) is close to $0.8E_1$. Therefore it appears natural to treat the
probability $X\sim \exp \left( -E_2/w\right) $ as the time-averaged fraction
of excited atoms in the condensate, and $(1-X)$, as the fraction of the
ground-state atoms. Such a condensate is a statistical mixture of ground-
and excited-state atoms at lattice sites. The excitations being pairwise,
this gives rise to the formation of virtual $(Xe)_2$ molecules in the
singlet state, which determine both the mean lattice bonding energy and the
equilibrium distances, because ground-state atoms are bound by attractive
forces only. The existence of the virtual $(Xe)_2$ molecules was deduced
from optical data \cite{15}. This scheme of atomic bonding can be described
by a periodic alternation of the bonding type (''resonance'').

If the $(Xe)_2$ molecule is considered as hydrogenlike, its binding energy
should be $\sim 1eV$. The average energy per Xe atom in the condensate $\sim
0.13eV$. Therefore $X\sim 0.13/1\sim \exp \left( -4r_2/5r_1\right) $. In
metallized $Xe$, $r_{2m}=1.47\AA $ ($r_2=2.2\AA $) and, hence, $X_m\sim 0.35$%
, which yields for the binding energy about $0.35(2.2/1.47)=0.5eV$. There
are many MCs with binding energies close to this value with no pressure
applied altogether (see below).

A pair of electrons in a $(Xe)_2$ molecule may be considered as a boson, or
as a Cooper pair two lattice constants in size and with a binding energy of $%
\sim 1eV$. Under compression, this energy increases $(2.2/1.47)$ times to
reach $1.5eV$. This figure is close to the above estimates of the gap
energy when treating metallic $Xe$ as a superconductor.

MCs present apparently the only possibility of following continuously the
process of boson condensation at equilibrium and of realizing consecutive
transitions from an excitonic insulator to a superconductor and to a Fermi
metal, where the average fermion energy exceeds the average boson
energy of the
system for $P\gg P_m$ \cite{16,17}. When forced close together, stable $Cs_2$
molecules or $Ba$ atoms will immediately transfer to a Fermi metal by
passing through the unstable superconducting phase (the Mott transition),
the situation occurring, for instance, when a metal is ''diluted'' by an
insulator \cite{6,15}. Steady-state $H_2$ molecules brought into proximity
produce virtual $(H_2)_2$ molecules, and it is the latter that account for
the properties of condensed hydrogen.

Compression of $Xe$ increases the concentration of $(Xe)_2$ molecules, and
they form chains and clusters, between which the electron tunneling sets in.
This may account for the frequency dependence of absorption in the
pretransitional pressure region \cite{10} (Fig. \ref{fig3}). The possibility
of existence, prior to Bose condensation, of such chains (``cycles''), which
nucleate through attraction between particles, was conjectured from
theoretical considerations \cite{18}.

The possibility of attaining superconductivity under compression can be
studied using MCs with complex molecules which do not possess magnetic
moments. Ground-state molecules with magnetic moments act as magnetic
impurities with a concentration $\sim (1-X)$, thus reducing dramatically $T_c
$. This situation is realized, for instance, in $O_2$ at metallization,
where $T_c=0.6K$ \cite{19}.

\begin{figure}
\epsfxsize=85mm
\epsffile{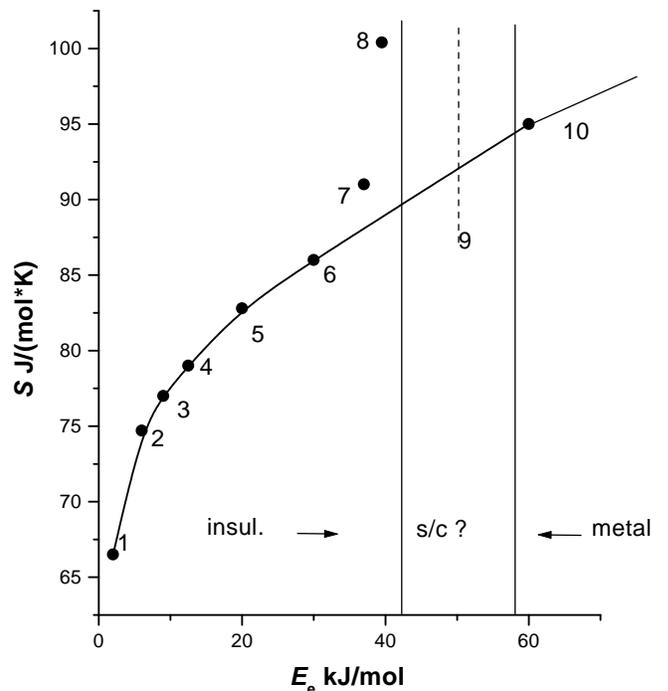}
\caption{
Dependence of the entropy of evaporation $S$ on binding energy $E_e$
for a number of molecular condensates and metals: (1) - $Ne$,
(2) - $Ar$, (3) - $Kr$, (4) - $Xe$, (5) - $CF_2Cl_2$, (6) - $C\,Cl_4$,
(7) - $C_6H_5Cl$, (8) - $H_2O$, (10) - $Hg$. Dashed line (9)
specifies the binding energy estimate for metallic xenon.
}
\label{fig4}
\end{figure}

Figure \ref{fig4} plots the entropy of evaporation $S$ as a function of
binding energy for MCs and metals. The energies of about $0.4eV$ for the
most typical MCs and $0.6eV$ for $Hg$ (metals) bound an empty region. When
''forced'' into metallization, the binding energy of $Xe$ ($\sim 0.5eV$)
falls into this instability region separating the insulators from metals
\cite{16,17}. In the case of complex molecules, the pressures may be lower,
but there is a danger of their pressure-induced fracture or ``chemical''
metallization.

The concentration of virtual molecules can probably be increased, and stable
superconductors obtained, without applying pressure at all. When an MC is
brought in contact with a metal, adsorption forces transfer adsorbed
molecules to the excimer state (the catalytic effect \cite{20}). The
fullerides $(Me_3)C_{60}$ belong possibly to such systems. In the
closely-packed array of the $C_{60}$ spheres $\sim 10\AA $ in diameter,
there are three voids per sphere, which contain one metal atom each.
Superconductivity sets in with $T_c$ of up to $\sim 40K$ \cite{21}. In
complex molecules, virtual excitations are created apparently in their
composite parts. It is such systems (clathrates), in which at least one of
the components is a molecular condensate, that are of practical interest.
Considered from this standpoint, in conventional HTSCs the system is
stabilized by ``chemical'' bonds, and the ``low'' $T_c$ temperatures are
determined by magnetic atoms.

---{\em Conclusions. }The properties of metallized xenon may thus far be a
manifestation of superconductivity with $T_c>300K$.

Molecular condensates are systems in which the Bose condensate state
(intermediate between the insulator and the metal) may be detected; this
state is unstable in atomic systems bound through stationary states (the
Mott transition).

\end{multicols}

\end{document}